\documentclass{article}

\usepackage[dvipsnames]{xcolor}
\usepackage{spconf,amsmath,graphicx}
\usepackage{tikz}
\usepackage{xspace}
\usepackage{hhline}
\usepackage{makecell}
\usepackage{soul}
\usepackage{array}
\usepackage{ifpdf}
\usepackage{textcomp}
\usepackage{siunitx}
\usepackage{graphicx}
\newcolumntype{L}[1]{>{\raggedright\arraybackslash}p{#1}}
\newcolumntype{C}[1]{>{\centering\arraybackslash}p{#1}}

\newcommand{\llcao}[1]{{\color{red} [llcao: #1]}}
\newcommand{\doutre}[1]{{\color{cyan} [doutre: #1]}}

\newcommand{\zhiyun}[1]{{\color{magenta} [zhiyun: #1]}}
\newcommand{\AR}[1]{{\color{violet} [arunnt: #1]}}
\newcommand{\CC}[1]{{\color{olive} [CC: #1]}}
\newcommand{\eat}[1]{}

\newlength\cwidth
\usetikzlibrary{calc,trees,positioning,arrows,chains,shapes.geometric,%
    decorations.pathreplacing,decorations.pathmorphing,shapes,%
    matrix,shapes.symbols}
\tikzset{
>=stealth',
  punktchain/.style={
    rectangle, 
    rounded corners, 
    draw=black, very thick,
    text width=10em, 
    minimum height=3em, 
    text centered, 
    on chain},
  line/.style={draw, thick, <-},
  element/.style={
    tape,
    top color=white,
    bottom color=blue!50!black!60!,
    minimum width=8em,
    draw=blue!40!black!90, very thick,
    text width=10em, 
    minimum height=3.5em, 
    text centered, 
    on chain},
  every join/.style={->, thick,shorten >=1pt},
}

\title{Improving Streaming Automatic Speech Recognition with Non-streaming Model Distillation on Unsupervised Data}

\name{\begin{tabular}{c}
      Thibault Doutre*, Wei Han*, Min Ma, Zhiyun Lu, Chung-Cheng Chiu, Ruoming Pang, \\
      Arun Narayanan, Ananya Misra, Yu Zhang, Liangliang Cao \thanks{$^*$The two authors contributed equally.}
      \end{tabular}}
\address{Google Inc., USA\\
\footnotesize{\texttt{\{doutre,weihan,llcao\}@google.com}}}
\begin{document}
\ninept
\maketitle

\begin{abstract}
Streaming end-to-end automatic speech recognition (ASR) models are widely used on smart speakers and on-device applications. Since these models are expected to transcribe speech with minimal latency, they are constrained to be causal with no future context, compared to their non-streaming counterparts. Consequently, streaming models usually perform worse than non-streaming models. We propose a novel and effective learning method by leveraging a non-streaming ASR model as a teacher to generate transcripts on an arbitrarily large data set, which is then used to distill knowledge into streaming ASR models.
This way, we  scale the training of streaming models to up to \emph{3 million hours} of YouTube audio. Experiments show that our approach can significantly reduce the word error rate (WER) of RNN-T models not only on LibriSpeech but also on YouTube data
in four languages. For example, in French, we are able to reduce the WER by 16.4\% relatively to a baseline streaming model by leveraging a non-streaming teacher model trained on the same amount of labeled data as the baseline.
\end{abstract}

\begin{keywords}
speech recognition, streaming ASR, non-streaming ASR, model distillation
\end{keywords}

\section{Introduction}
\label{sec:intro}

The advent of smart speakers such as Google Assistant, Siri, and Alexa has motivated a new generation of on-device recognition systems. End-to-end streaming models~\cite{graves2012sequence, he2019streaming, yeh2019transformer, zhang2020transformer, li2020towards, moritz2020streaming, tsunoo2019towards} have become attractive for on-device recognition tasks in two aspects: first, end-to-end models are usually compact, which makes them suitable to be used on devices. Second, such models often have a low latency (i.e. streaming), which is crucial to facilitate human-computer interactions -- an automated assistant can only engage the user when it responds quickly to requests. 

Contrary to non-streaming ASR models such as Chorowski et al's attention-based models~\cite{chorowski2015attention} or Chan et al's listen-attend-spell models \cite{chan2016listen}, streaming ASR models cannot utilize the full context. In the past few years, many research efforts have been devoted to improving streaming ASR~\cite{sainath2019two,sainath2020streaming,Saon-distillation-2020}. 
However, a key question that remains is how to utilize unlabeled data, especially for non-English languages with much less training data. 

In~\cite{liao2013large}, Liao et al showed that we could generate large-scale training data from the public YouTube videos, leveraging transcripts uploaded by the video owners. Their method ~\cite{liao2013large} is called ``Island of Confidence" because it identifies segments of audio that have correct transcripts with high confidence. In this paper, we name the data generated by \cite{liao2013large} as \textit{Confisland} for short.
Because of the continuously increasing amount of YouTube data with user-uploaded transcripts, such \textit{Confisland} data is a good resource to train end-to-end ASR models. 


In this paper, we propose a new approach to train end-to-end streaming models from unsupervised data. Our approach can be divided into three steps: (1) We employ the state-of-the-art full-context model as a teacher model. 
(2) We convert unlabeled audio sequences into random segments and transcribe them using the full-context teacher model. (3) We use the waveforms and their predicted transcripts in a noisy student learning framework~\cite{park2020improved,xie2020self,he2019revisiting,li2017learning}. 
Our method can potentially collect much more data than \textit{Confisland} \cite{liao2013large} as it can also use audio data without user-uploaded transcripts. Besides, since full-context ASR models perform significantly better than streaming models, they work as stronger teachers and finally foster more robust streaming student models.  \cite{Saon-distillation-2020} also proposed to use a full-context model as the teacher for RNN-T models, but their approach uses posteriors as targets and therefore requires additional distillation pretraining to address the alignment mismatch between the student and the teacher. Our approach is more efficient as we use the predicted transcripts as targets directly which does not have the alignment issue and requires only one distillation step.

\section{Method}
In this section, we describe our recipe to improve the performance of streaming end-to-end ASR models. We first introduce streaming and non-streaming models and then present our teacher-student training framework. 

\subsection{Streaming and non-streaming end-to-end ASR models}
\label{ssec:streaming}
Streaming end-to-end models~\cite{li2020towards} produce and update hypotheses frame-by-frame. For example, CTC or RNN-T models with unidirectional encoders fall into this category. They are popular candidates for on-device speech recognition due to their low latency and small memory footprint. However, 
streaming models usually perform worse than non-streaming models.

In this paper, we focus on improving the performance of a streaming RNN-T model~\cite{narayanan2019recognizing}.
The model has an encoder network of 8 layers of unidirectional LSTMs with 2048 cells. Each LSTM layer has a projection layer of 640 outputs for parameter efficiency. The decoder consists of 2 unidirectional LSTMs, also with 2048 units and 640 projections similar to the encoder layers. The joint network is a fully connected layer with 640 units. The target is a
sequence of word piece tokens~\cite{schuster2012japanese,narayanan2019recognizing,chiu2020rnn}.
This makes up a total of 122 million parameters. The front end is 128-channel filter banks, computed from a 32ms window with a stride of 10ms. During training, the utterances are re-sampled to generate both 8k and 16k sampling-rate versions for robustness. For SpecAugment, we use mask parameter $F$ of 27, twice masks with the time-mask ratio of 1.0. The RNN-T model is trained with the regularization technique introduced in~\cite{chiu2020rnn}, i.e. variational weight noise, SpecAugment~\cite{park2019specaugment}, and random state sampling and random state passing~\cite{narayanan2019recognizing}. 


Unlike streaming models, non-streaming models~\cite{chorowski2015attention,chan2016listen} examine all of the speech input before producing output hypotheses. For example, RNN-Ts with bi-directional encoders and attention-based models can use full context to achieve lower WERs. But they suffer from high latency, which is not suitable in applications like smart assistants. This paper considers two non-streaming transducer models: the Conformer model~\cite{gulati2020conformer} and the TDNN model~\cite{chiu2020rnn}.

The Conformer transducer model has shown substantial improvements over RNN-T models in speech recognition. It takes advantage of both convolution neural networks (CNN) and transformers' architectures in the encoder to capture the local and global context in the audio~\cite{gulati2020conformer}. We use a model with 16 conformer blocks in the encoder and 1 LSTM decoder layer which has 2048 cells with a projection layer of 640 outputs. This makes up a total of 179 million parameters. Note that the attention layer encodes all frames in the utterance simultaneously, and is thus non-streaming.

The TDNN model stacks 3 macro layers in the encoder. Each macro layer consists of a 1-D convolution, a 1-D max pooling, and 3 bi-directional LSTM layers with 512 hidden units  in each direction and a 1536-dimensional projection~\cite{pang2018compression}. The decoder network has a uni-directional LSTM with 1024 hidden units. The joint network has 512 hidden units and the final output uses a 4k word piece model. For both Conformer and TDNN  models, we follow~\cite{gulati2020conformer} to set the frontend and SpecAugment hyper-parameters.

\subsection{Teacher-student training}
\label{ssec:approach}
\begin{figure}[tpb]
\begin{tikzpicture}
  [node distance=.5cm,
  start chain=going below,]
  \node[punktchain, on chain=going right] (yt) {Unlabeled Youtube};
  \node[punktchain, join] (segments) {Random Segments};
  \begin{scope}[start branch=hoejre]
    \node(conformer) [punktchain, on chain=going left, fill=gray!10] {Non-Streaming Teacher};
  \end{scope}
  \node[punktchain, join,] (preds) {Teacher's Predictions};
  \node[punktchain, join,] (noise) {Add Noise + SpecAugment};
  \node[punktchain, join, fill=gray!10] (student) {Streaming Student};
  \draw[|-,-|,->, thick,] (conformer.south) |-+(0,-0.5em)-| (preds.north);
\end{tikzpicture}
\caption{Our method trains a streaming model, learning from the predictions of a powerful non-streaming teacher model on large-scale unlabeled data via a teacher-student training framework. See Sect.~\ref{ssec:approach} for more details.}
\label{fig:diagram}
\end{figure}
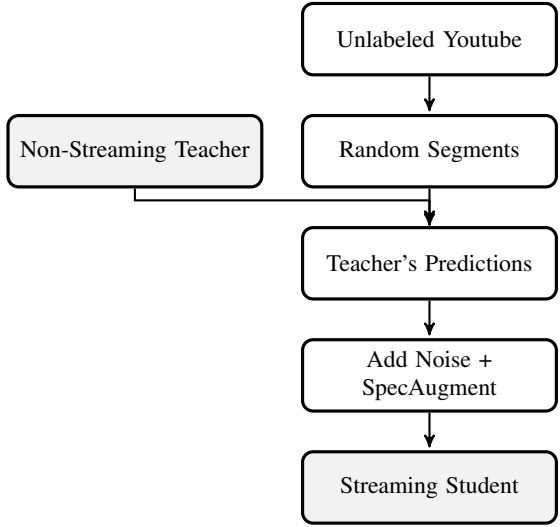

One big challenge for training end-to-end models is that they are notoriously data-hungry. A straightforward approach to solve the data challenge is to borrow the method described in \cite{liao2013large} which can generate a lot of training samples from YouTube. However, such \textit{Confisland} data requires that the YouTube audio must be associated with user-generated transcripts so that the model can align audio with text and select the most confident samples for training. In this paper, we use a simpler yet more effective approach named teacher-student training framework~\cite{park2020improved,xie2020self,pang2018compression} to collect pseudo-labels~\cite{lee2013pseudo} on unlabeled audio sequences.

Fig.~\ref{fig:diagram} illustrates our idea. Given unlabeled YouTube data, we convert them into segments with random lengths. Such \textit{random segments} do not require an alignment model while still providing good samples for ASR training. A non-streaming teacher model is then used to transcribe these random segments. These predictions can be viewed as pseudo labels. The random segments with these pseudo labels will be used to train a student model. Following \cite{park2020improved}, we augment the inputs with noise when training the student network, to make the student model more robust. 

Our method can be viewed as an extension of the noisy student learning in \cite{park2020improved} with the following novelties: (1) We use a non-streaming model as a teacher and a streaming model as a student, while the work in  \cite{park2020improved} uses the same model. (2) We find that unsupervised \textit{random segments} can be as good  or even better than \textit{Confisland} data. (3) This approach is very scalable. This paper manages to train a streaming model on 3 million hours of YouTube (unlabeled) data, orders of magnitude larger than typical supervised ASR training sets, and significantly improves WERs in four languages. The experimental section will explain these discoveries in more detail.

\section{Experiments on LibriSpeech}
We first validate our method on the public LibriSpeech 960 hour dataset \cite{panayotov2015librispeech}. 
One important factor for assessing streaming end-to-end ASR models on Librispeech is to have consistent latency metrics and criteria. Previous works~ \cite{moritz2019triggered, wang2020low, moritz2020streaming,tsunoo2020streaming, univ_asr2020} use different latency metrics and criteria which makes it difficult to directly compare results across them. Our task uses the same latency metric and criterion as \cite{univ_asr2020}, as well as the same streaming model as described in \cite{univ_asr2020} to verify whether it can benefit from learning from non-streaming teachers. We use the non-streaming Conformer as the teacher model to transcribe the unlabeled 60K subset of LibriLight \cite{librilight}. Then, we use both LibriSpeech and LibriLight to train another student streaming model using the same structure as \cite{univ_asr2020}. Table~\ref{tab:librispeech} shows that the WER of the student model improves to 3.3/8.1 on the \textit{test-clean} and \textit{test-other} sets, respectively. 

This experiment is motivated by Park et al's work on noisy student learning \cite{park2020improved}. Our work differs from \cite{park2020improved} in two aspects. First, all models in \cite{park2020improved} are non-streaming models. Second, our experiment on LibriSpeech does not use language model fusion or any data filtering. This last point ensures consistency with future experiments in this work: we do not have good language models in languages other than English, and 
the filtering step does not help large scale data like YouTube. 
The simplified experiment on Librispeech validates the effectiveness of leveraging the predictions of a non-streaming teacher on unlabeled data and motivates us to apply this method on YouTube data in Sect~\ref{sec:exp}. 

\begin{table}[hbtp]
\caption{WERs of different models on LibriSpeech. The streaming baseline model and the non-streaming teacher are trained on LibriSpeech 960h. The streaming student model is trained on both LibriSpeech 960h and the predictions of the non-streaming teacher on LibriLight.}
\centering
\begin{tabular}{ |p{0.22\cwidth}|C{0.25\cwidth}|C{0.28\cwidth}|C{0.25\cwidth}| }
    \hline
    & Streaming baseline \cite{univ_asr2020} \eat{model on Librispeech 960h} & Non-streaming teacher  \eat{on Libri-light} & Streaming student  \eat{on Libri-light} \\ 
    \hline
    test-clean & 4.6  & 1.7  & 3.3 \\
    test-other & 9.7 &  3.8 & 8.1 \\
    \hline
\end{tabular}
\label{tab:librispeech}
\end{table}

\section{Experiments on YouTube data}
\label{sec:exp}

\subsection{Evaluation sets}
\label{ssec:datasets}
This paper considers end-to-end ASR models in four different languages: French, Spanish, Portuguese, and Italian. 
To benchmark the performance of speech recognition, we first use the publicly available \textit{Common Voice} data with transcribed short form utterances in 29 languages \cite{CommonVoicePaper}. However, utterances in \textit{Common Voice} are usually much shorter than those in \textit{Confisland} and limited in terms of the diversity of the content. So we also introduce another test set from YouTube, \textit{YT-long}, in which the utterances lengths vary between 40 seconds and 30 minutes. \textit{YT-long} was generated by sampling and hand-transcribing popular videos from YouTube based on view counts. Note that videos in \textit{YT-long} are much longer than those in the training samples: they present a very challenging test set for end-to-end ASR models trained on small training sets. Also, the nature of \textit{Common Voice} data is different from YouTube data and can be considered out-of-domain in this work. 

\subsection{Models trained from Confisland data}

Collecting transcribed speech data in various languages can be very expensive. Liao et al's work \cite{liao2013large} enabled us to collect semi-supervised data from YouTube. The \textit{Confisland} data set is built using transcripts uploaded from YouTube users. Following \cite{liao2013large}, we gathered audio data in different languages. Note that the non-English \textit{Confisland} data sets are much smaller than the English one, mainly because there are fewer user-uploaded transcripts for non-English videos. For example, there are more than 200K hours of audio from the English \textit{Confisland} data set, but in Spanish, there are at most 12K hours. The other languages (French, Portuguese, and Italian) have even less audio data from \textit{Confisland}. Consequently, RNN-T models trained from non-English \textit{Confisland} data do not perform as well as in English. 
Table~\ref{tab:languagesteacher} summarizes the WERs of these models. It is easy to see that the streaming models (RNN-T) are consistently worse than the non-streaming models in all four languages. For example, the non-streaming models' WER on Portuguese reduced by relative 25.9\% on \textit{YT-long} (22.8\% vs. 30.8\%), and relative 16.5\% on \textit{Common Voice} (25.8\% vs 30.6\%). 

\begin{table}[h]
\centering
\caption{WERs of ASR models trained on \textit{Confisland}.}
\begin{tabular}{ |p{0.2\cwidth}|p{0.3\cwidth}|C{0.2\cwidth}|C{0.3\cwidth}| }
    \hline
     & Test set & Streaming model on \textit{Confisland} & Non-streaming teacher model on \textit{Confisland} \\
    \hline
    French & YT-long &34.5 & 18.6  \\
     & Common Voice &36.2 & 33.2 \\
    \hline
    Spanish & YT-long &35.9 & 18.6  \\
     & Common Voice &22.0 & 11.2  \\
    \hline
    Portuguese & YT-long &30.8 & 22.8 \\
     & Common Voice &30.9 & 25.8 \\
    \hline
    Italian & YT-long &24.0 & 16.2 \\
     & Common Voice &30.0 &27.3 \\
    \hline
\end{tabular}
\label{tab:languagesteacher}
\end{table}


\subsection{Our approach using random YouTube segments}
\label{ssec:results}

To improve the performance of streaming models, we apply our method presented in Section~2.
Note that our method can utilize any unsupervised audio. However, we first report results using only the original set of audio used to generate \textit{Confisland}. 
We use the same list of audio sequences from YouTube, and then randomly cut audio into segments with lengths varying between 5 seconds and 15 seconds. To make the pipeline simple to use in various scenarios, we choose not to use complicated segmentation methods other than random segmentation. We found such a simple method works better than fixed-length segmentation. We call this segmented unlabeled set \textit{YT-segments}. Note that the total number of hours in \textit{YT-segments} is greater than the number of hours in \textit{Confisland}, since the latter uses additional filtering strategies \cite{liao2013large}. The size of the training data in each language is summarized in Table \ref{tab:data}. 

\begin{table}[t]
\caption{Number of hours of the \textit{Confisland} and \textit{YT-segments} data sets, for different languages. Data in \textit{YT-segments} are generated by randomly segmenting the original YouTube videos used by \textit{Confisland} (pre-filtering).}
\begin{tabular}{ |p{0.35\cwidth}|C{0.35\cwidth}|C{0.35\cwidth}| } 
\hline
 & \textit{Confisland} & \textit{YT-segments} \\ 
\hline
French& 10,353  & 24,405 \\ 
Spanish & 13,468  & 34,762   \\ 
Portugese & 1,660 & 2,876   \\ 
Italian & 6,742 & 13,093  \\ 
\hline
\end{tabular}
\label{tab:data}
\end{table}

To utilize the unlabeled segments, we choose the non-streaming model in 
Table~\ref{tab:languagesteacher} as the teacher model to predict the transcripts of \textit{YT-segments}. Note that we can choose other models as the teacher model, or use other training sets to train the teacher. The WERs of teacher models can be found in Table \ref{tab:languagesteacher}. 

Finally, we train a streaming RNN-T model using the teacher's predictions. Table \ref{tab:languages} reports the WERs of baseline and student models. By leveraging the same amount of labeled data (\textit{Confisland}), student models constantly outperform baseline models on \textit{YT-long}. For example, the absolute WER improved from our baseline by 9.5\% in French, 7.9\% in Spanish, 2.5\% in Portuguese, and 3.2\% in Italian. As for \textit{Common Voice}, the WERs also improved: by 1.5\% in French, 5.5\% in Spanish, 2.0\% in Portuguese, and 6.4\% in Italian. 
This suggests that learning from the teacher's predictions on random segments is more effective than learning from \textit{Confisland} data.

\begin{table}[b]
\caption{Comparing the WERs of streaming RNN-T models trained on \textit{Confisland}
with the model from our distillation approach trained on the corresponding random segments.
}
\begin{tabular}{ |p{0.2\cwidth}|p{0.3\cwidth}|C{0.2\cwidth}|C{0.3\cwidth}| }
    \hline
     & Test set & 
     Streaming model on \textit{Confisland} & 
     Streaming student on \textit{YT-segments} \\ 
    \hline
    French & YT-long & 34.5 & 25.0 \\
     & Common Voice & 36.2 & 34.7 \\
    \hline
    Spanish & YT-long & 35.9 & 28.0  \\
     & Common Voice & 22.0 & 16.5 \\
    \hline
    Portuguese & YT-long & 30.8& 28.3 \\
     & Common Voice & 30.9 & 28.9  \\
    \hline
    Italian & YT-long & 24.0 & 20.8 \\
     & Common Voice & 30.0 & 23.6 \\
    \hline
\end{tabular}
\label{tab:languages}
\end{table}

\begin{figure}[t]
\includegraphics[width=8.5cm]{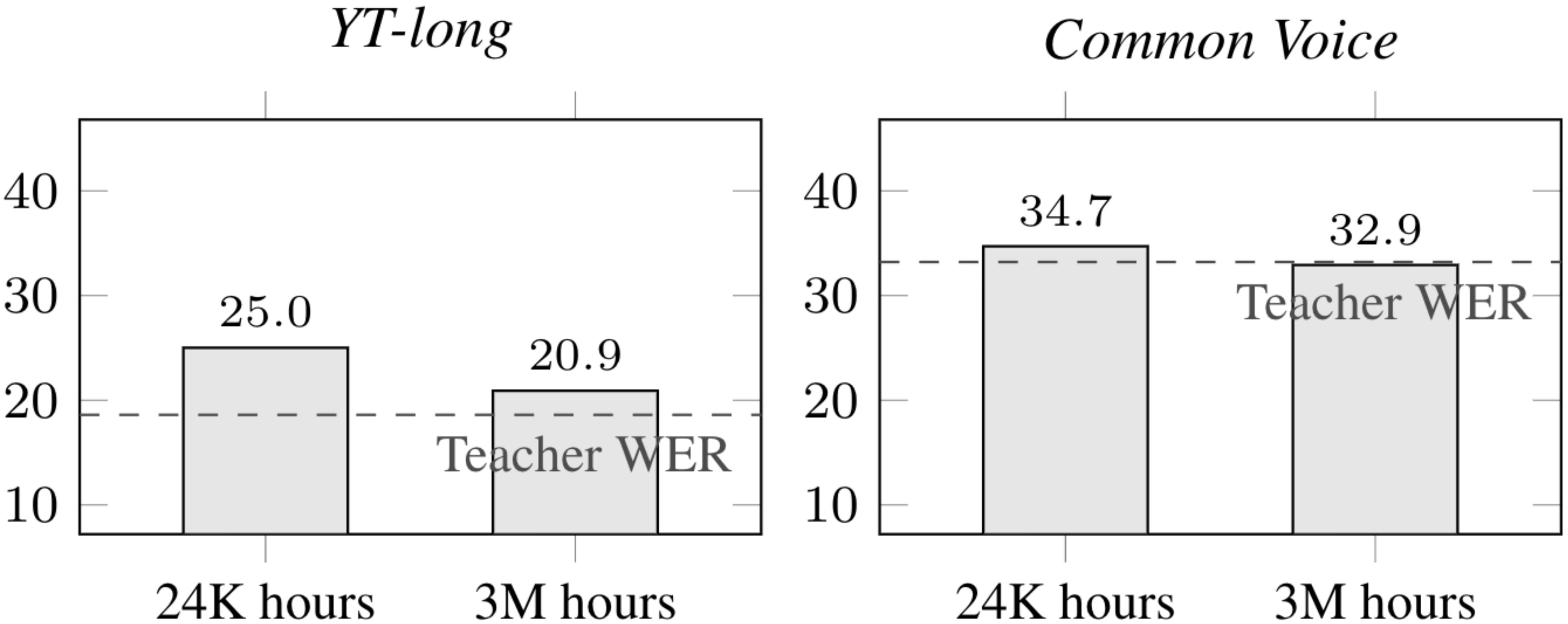}
\vspace{-0.3em}
\caption{How scaling the unlabeled data set \textit{YT-segments} impacts the student model's WER. Utterances are transcribed using the same Conformer model.}
\label{fig:barplotytbig}
\end{figure}

Note that Table~\ref{tab:languages} only compares our method using the same audio list as in \textit{Confisland}. But in practice, our method can be further improved by leveraging more unlabeled data. Since our method only requires a teacher model but does not require user uploaded transcripts, we consider scaling the number of utterances in \textit{YT-segments}. Take French data as an example: by scaling up the training data set, we hope that the performance of the student model (25.0\% WER) would improve to eventually get closer to the performance of the teacher model (18.6\% WER). \textit{3 million hours} of French audio are gathered from YouTube and then randomly segmented into utterances of lengths varying between 5s and 15s. This new data set has over 1 billion utterances and is 125 times larger than the original \textit{YT-segments}. The results are reported in Fig. \ref{fig:barplotytbig}. The WER on \textit{YT-long} drops significantly, from 25.0\% to 20.9\%. The WER of our out-of-domain set \textit{Common Voice} also improves, from 34.7\% to 32.9\%.

\subsection{Ablation studies}
\label{ssec:ablationstudy}

In this section, we explore how the different components of our method affect the performance of student models. We focus on the influence of different teachers and the lengths of random segments. 

\subsubsection{Training from different teachers} 
\label{ssec:teachers}

\begin{figure}[t]
\includegraphics[width=8.5cm]{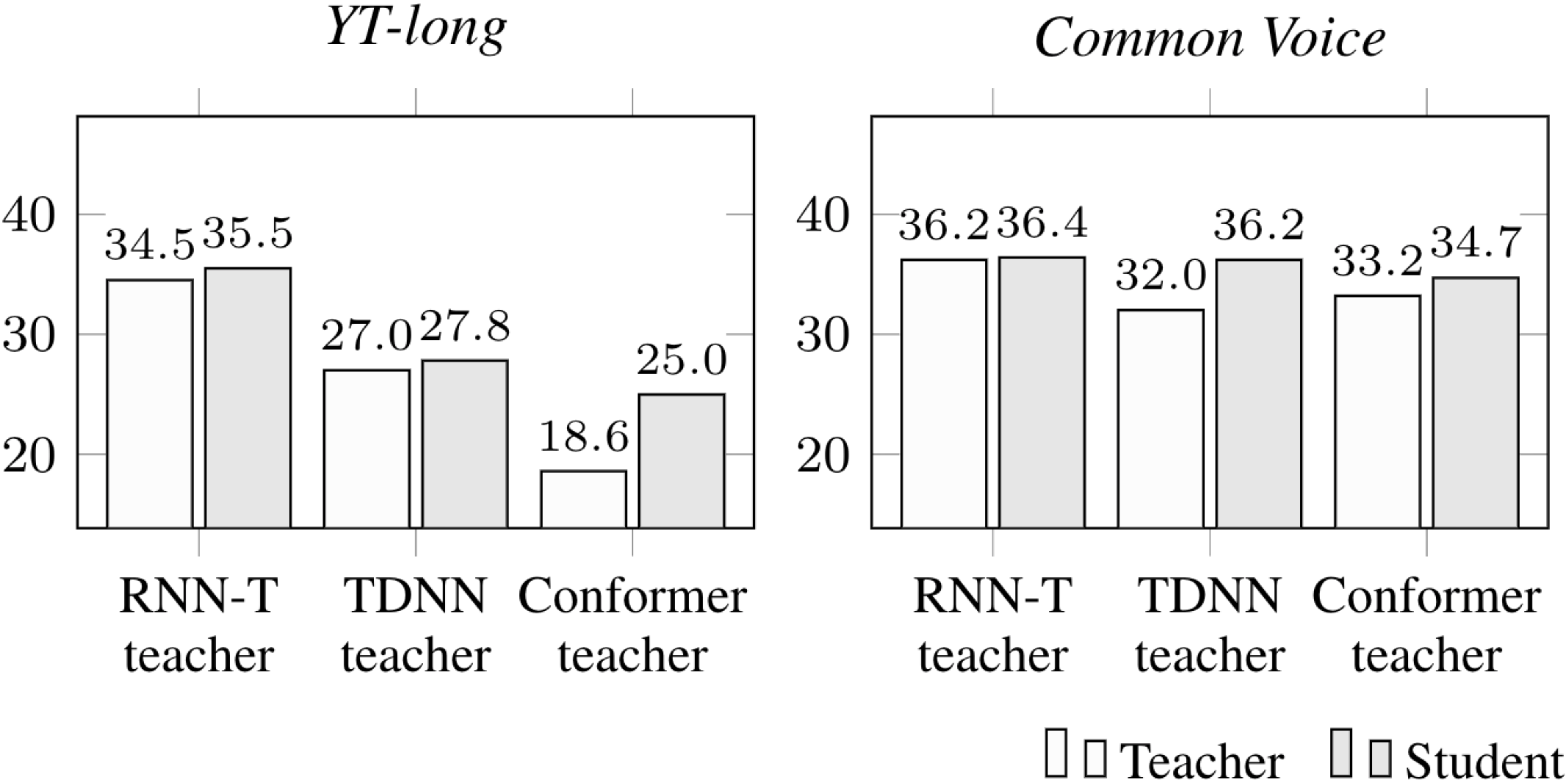}
\vspace{-0.3em}
\caption{WERs of teacher models and their respective RNN-T students on French data. Teacher models are trained on \textit{Confisland} and student models use the predictions of their respective teachers on \textit{YT-segments}.}
\label{fig:barplotteachers}
\end{figure}

Our final student model is trained from a teacher's predictions.
Therefore, it is intuitive to think that better teachers lead to better students. 
We aim to provide evidence of this claim by looking at two different teacher models: the non-streaming TDNN \cite{chiu2020rnn} and the non-streaming Conformer \cite{gulati2020conformer}. Both non-streaming models are trained on \textit{Confisland}. Results are summarized in Fig. \ref{fig:barplotteachers}. The Conformer teacher has the lowest WER on \textit{YT-long}, and the student trained from its predictions also has the lowest WER among students. When using the same streaming RNN-T model as a teacher and as a student, we see that the performance degrades. Indeed, the RNN-T teacher model has a higher WER on \textit{YT-long} (34.5\%) compared to the non-streaming teachers: 27.0\% and 18.6\% on \textit{YT-long}. A higher WER for the teacher also translates into a higher WER for its streaming student (see Fig. \ref{fig:barplotteachers} for details). The same trend is observed on \textit{Common Voice}. We conclude that our teacher-student framework works best when using a strong, non-streaming teacher.

\subsubsection{Lengths of YT-segments} 
\label{ssec:lengths}
\begin{figure}[t]
\includegraphics[width=8.5cm]{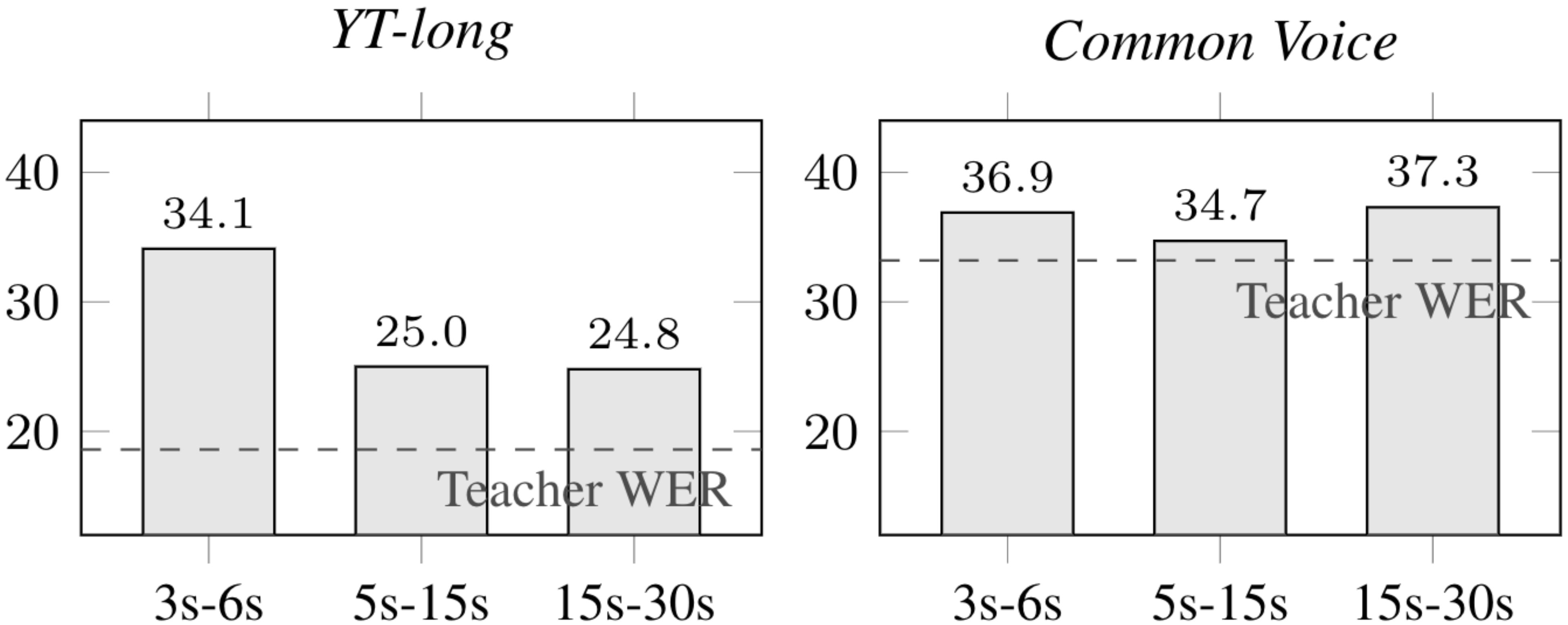}
\vspace{-0.3em}
\caption{How the segmentation of \textit{YT-segments} impacts the WER of the student model. \textit{YT-segments} are transcribed using the same Conformer teacher model. Only the lengths of utterances vary, not the total number of hours in each version of \textit{YT-segments}.}
\label{fig:barplotsegmentation}
\end{figure}

Segmenting utterances from YouTube to get \textit{YT-segments} can be done in numerous ways. We explore training student models from utterances with different segmentation lengths. Three versions of \textit{YT-segments} are generated, using the same number of hours of audio. Audio sequences are randomly split into utterances of respective lengths of 3s to 6s, 5s to 15s, and 15s to 30s. 
With the same Conformer teacher,
Fig. \ref{fig:barplotsegmentation} compares the performance of RNN-T student models trained from these different versions of \textit{YT-segments}.  
We notice that training from shorter utterances harms the performance of the student on \textit{YT-long}. Training from utterances of 15s to 30s doesn't seem to help much on \textit{YT-long}, and the error on \textit{Common Voice} increases.

\section{Conclusion}
\label{sec:conclusion}
In this paper, we proposed a teacher-student training framework to improve the performance of streaming end-to-end ASR models. The improvement comes from a powerful non-streaming teacher, as well as a large amount of unlabeled data. Our approach consistently improved streaming ASR models trained on Librispeech and Youtube data. On Youtube French data, we reduced the WER from 34.5\% to 20.9\%, a 39.4\% relative improvement, by training on 3 million hours of unlabeled audio. We found the unsupervised random segments more effective than \textit{Confisland} data from YouTube in French, Spanish, Portuguese, and Italian.
In the future, we plan to explore more effective learning methods and also extend the large scale unlabeled learning to more languages.

\section{Acknowledgement}
We are very thankful for our colleagues Basi García, Jiahui Yu, Bo Li, Hank Liao, Yonghui Wu, Françoise Beaufays, and Trevor Strohman for their help and suggestions to improve this work. 

\vfill\pagebreak

\bibliographystyle{IEEEbib}
\bibliography{bibliography}

\end{document}